\documentclass[screen, authorversion, nonacm]{acmart}
\AtBeginDocument{%
  \providecommand\BibTeX{{%
    \normalfont B\kern-0.5em{\scshape i\kern-0.25em b}\kern-0.8em\TeX}}}

\begin{document}

\title[Questionnaires for Everyone]{Questionnaires for Everyone: Streamlining Cross-Cultural Questionnaire Adaptation with GPT-Based Translation Quality Evaluation}

\author{Otso Haavisto}
\orcid{0009-0001-5865-1047}
\email{otso.haavisto@aalto.fi}
\affiliation{%
  \institution{Aalto University}
  \city{Espoo}
  \postcode{02150}
  \country{Finland}
}

\author{Robin Welsch}
\orcid{0000-0002-7255-7890}
\email{robin.welsch@aalto.fi}
\affiliation{%
  \institution{Aalto University}
  \city{Espoo}
  \postcode{02150}
  \country{Finland}}

\begin{abstract}
Adapting questionnaires to new languages is a resource-intensive process often requiring the hiring of multiple independent translators, which limits the ability of researchers to conduct cross-cultural research and effectively creates inequalities in research and society. This work presents a prototype tool that can expedite the questionnaire translation process. The tool incorporates forward-backward translation using DeepL alongside GPT-4-generated translation quality evaluations and improvement suggestions.
We conducted two online studies in which participants translated questionnaires from English to either German (Study 1; $n=10$) or Portuguese (Study 2; $n=20$) using our prototype. To evaluate the quality of the translations created using the tool, evaluation scores between conventionally translated and tool-supported versions were compared. 
Our results indicate that integrating LLM-generated translation quality evaluations and suggestions for improvement can help users independently attain results similar to those provided by conventional, non-NLP-supported translation methods. This is the first step towards more equitable questionnaire-based research, powered by AI.
\end{abstract}

\begin{teaserfigure}
  \includegraphics[width=\textwidth]{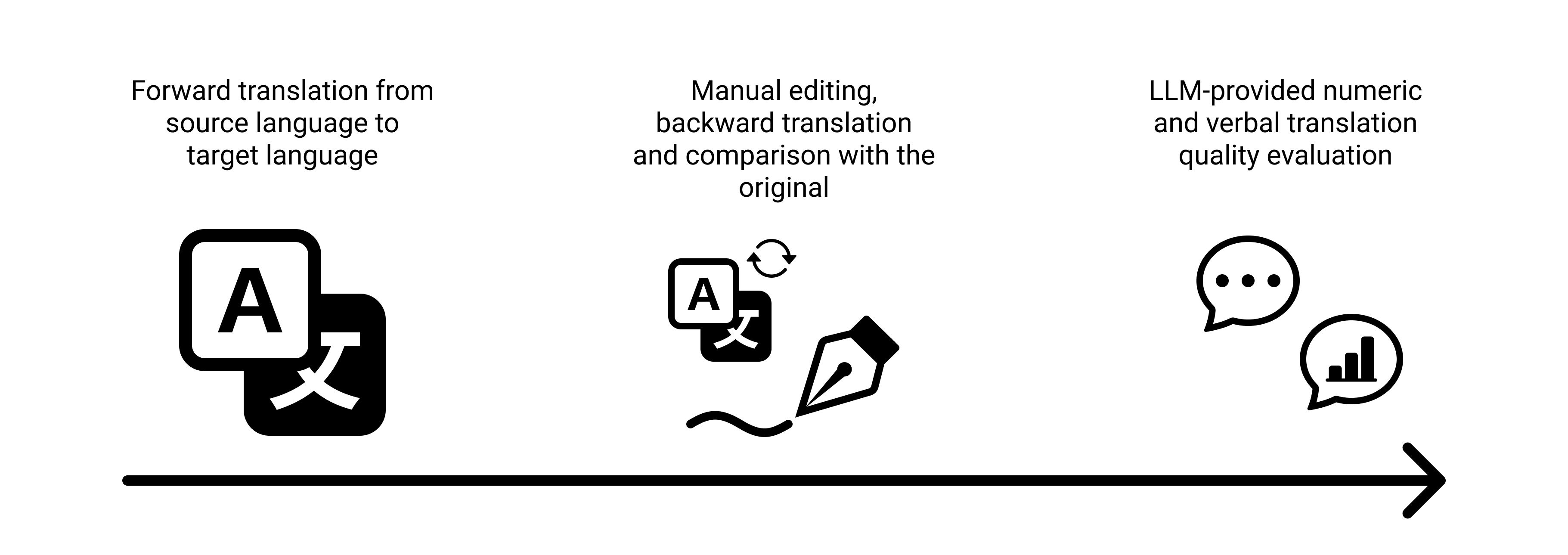}
  \caption{The translation flow of the final prototype application. Users are able to translate questionnaires forwards and backwards; edit the translations manually; and to request GPT-4-generated evaluation of their translation.}
  \Description{The flow within the prototype application. First, users enter the questionnaire to translate. After a forward machine translation to the target language has been performed, users can manually edit the translation and backtranslate it to the source language for comparison to the original. Finally, users can request a GPT-generated numeric and verbal evaluation of the translated questionnaire.}
  \label{fig:teaser}
\end{teaserfigure}

\maketitle

\section{Introduction}

Currently, most HCI studies utilizing standardized user experience questionnaires are conducted in Europe \cite{diaz-oreiro_ux_2021}. In psychology, the availability of clinical questionnaires vary greatly between high- and low-income countries \cite{ali_validated_2016, hoffman_mapping_2022}. This creates structural inequalities for conducting high-quality questionnaire-based research in non-English-speaking and low-income parts of the world. 

When questionnaires are adapted to new languages, Brislin’s \cite{brislin_back-translation_1970} translation methods for cross-cultural research are commonly cited. However, using these is not always straightforward. Cha et al. \cite{cha_translation_2007} note that, while Brislin’s backtranslation and committee methods undoubtedly yield the best results, the number of translators required to reach a consensus is not always known, which makes the process resource-intensive. For example, in the backtranslation approach, the source text is translated forward (to the target language) and backward (to the original language) by \textit{separate and independent} translators until the two versions are \textit{identical}.

In practice, this can be seen in the adoption of mixed-methods translation processes. Two recent studies adapting questionnaires to Finnish by Heilala et al. \cite{heilala_finnish_2023} and Westerlund et al. \cite{westerlund_finnish_2023} used combined approaches, with the authors taking part in the conflict resolution process in both cases and even doing a part of the translation in the latter. Notably, while both studies employed the backtranslation method, neither used it iteratively as intended \cite{brislin_back-translation_1970}. Partly automating the translation process would lower the threshold of iterative questionnaire translation, potentially resulting in higher-quality translations overall.

We set out to develop a prototype of a questionnaire translation tool that would exploit the versatility of large language models (LLMs) in natural language processing tasks (e.g. \cite{gillioz_overview_2020, floridi_gpt-3_2020, brin_comparing_2023, kasneci_chatgpt_2023}) to the benefit of researchers conducting cross-cultural studies. The prototype allows users to translate questionnaires, edit translations, backtranslate to the source language for comparisons against the original and receive evaluations of the quality of the translation generated by GPT-4 \cite{openai_gpt-4_2023}. 

To evaluate the assistive potential of LLMs in questionnaire translation, we conducted two online studies in which participants used our prototype to translate the ATI scale \cite{franke_personal_2019} from English to German (Study 1) and BFI-10 \cite{rammstedt_measuring_2007} from English to Portuguese (Study 2). In both studies, participants reached equivalent translation quality to conventionally translated (i.e. not translated using NLP tools) versions of the questionnaires using machine translation and GPT-4-generated quality evaluations and suggestions for improvement. Participants found the evaluations helpful and made use of the suggestions provided by the prototype: In the latter study, over half of the participants reported having implemented at least some of the suggestions.

\section{Background}

Cha et al. \cite{cha_translation_2007} note that there is \textit{“no gold standard”} for questionnaire adaptation due to varying resources and research contexts. Another literature review of guidelines for cross-cultural adaptation of questionnaires by Epstein et al. \cite{epstein_review_2015} could not find a consensus regarding recommendations either. However, while the translation methods themselves have not been standardized, the questionnaire adaptation process should address a few key issues.

Translation, in general, can be approached from multiple different angles, which in turn determine the criteria for a translation being \textit{good} \cite{house_translation_2014}. In questionnaire translation, the semantic similarity between the translated version and the original is often more important than word-by-word match \cite{su_generating_2002}; both versions should equally measure the same phenomena. 

The construction of the original questionnaire also affects how well it translates over to other languages: Using colloquial or culture-specific terminology makes it difficult for a translated questionnaire to be accurate \cite{brislin_back-translation_1970, su_generating_2002}. Finally, questionnaires should undergo an intensive validation process with real participants before they are deployed in research \cite{brislin_back-translation_1970, epstein_review_2015, taherdoost_validity_2016}.
 Kocmi and Federmann \cite{kocmi_large_2023} evaluated a number of LLMs as translation quality assessors. In their work, they developed a series of prompts that would return quantifiable scores for a translation, collectively named GEMBA. The authors argue that LLMs {\textemdash} especially GPT-4 {\textemdash} are \textit{“the state-of-the-art”} in translation quality assessment. Promising results were also achieved by Lu et al. \cite{lu_error_2023}, who attained a similar level of quality when prompting a verbal analysis of errors in a given translation when compared to a human reference.

The cross-cultural questionnaire adaptation process could, then, be considered to consist of two major steps: translation and validation. As all questionnaire translations should undergo a rigorous validation process before they are applied in practice regardless of the method used to translate them, incorporating LLM-prompted heuristic assessments of semantic similarity and suggestions for improvements to translation software has a higher potential to be helpful to researchers during the translation process.

\section{Prototype}
\subsection{Task Analysis}

To begin the prototyping process, hierarchical task analyses (HTA; \cite{shepherd_hta_1998}) were performed on existing methods for cross-cultural questionnaire adaptation. The methods selected were Brislin’s \cite{brislin_back-translation_1970} backtranslation and committee approaches and the recent approaches taken by Heilala et al. \cite{heilala_finnish_2023} and Westerlund et al. \cite{westerlund_finnish_2023} when translating the ATI \cite{franke_personal_2019} and CLEFT-Q \cite{tsangaris_establishing_2017} questionnaires into Finnish. The four HTA graphs are presented in Appendix \ref{app:hta} alongside snapshots of the prototyping process. Questionnaire validation was not broken down following the recommendations by both Cha et al. \cite{cha_translation_2007} and Epstein et al. \cite{epstein_review_2015} to keep the validation process separate.

The HTA identified three common features in the translation processes. First, all used independent translators at some point in the process. Second, all groups used discussions as a tool for assessing translation quality and resolving potential conflicts of opinion or ambiguities in meaning. Finally, three out of four of the processes described the use of forward-backward translation, in which a translated version of the questionnaire is translated back to its original language and the two original-language versions compared. The aim of using forward-backward translation is to assert a high level of interchangeability between the original and final translation.

\subsection{Prototype Construction}

An initial interactive wireframe based on the task analyses and sketches was constructed for usability testing. Two participants, both acquainted with HCI research (a doctoral researcher and a research assistant), were asked to translate a four-item questionnaire from English to Finnish while explaining their thought process. The issues uncovered by the usability tests were addressed during the development of the functional prototype.

\begin{figure}
    \centering
    \includegraphics[width=1\linewidth]{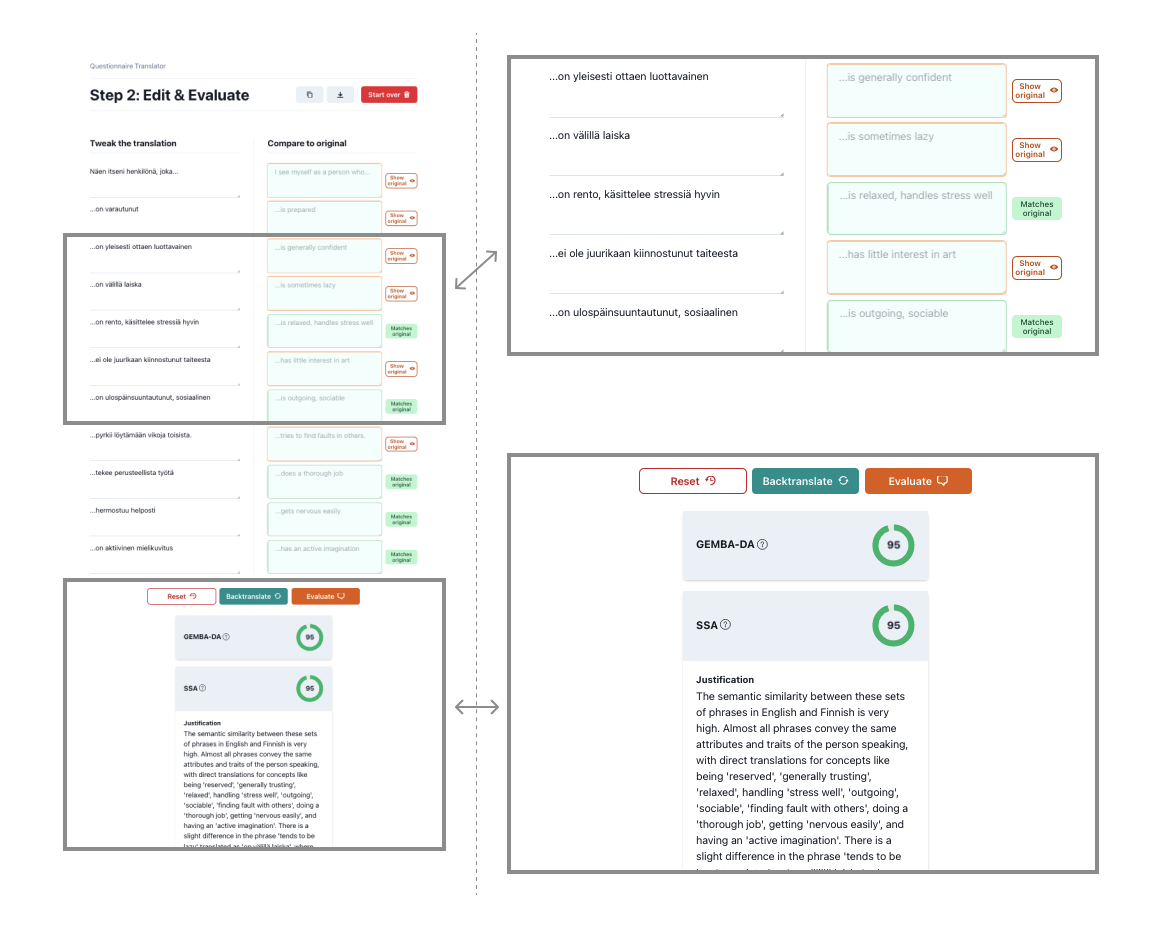}
    \caption{The view for editing and evaluating translations in the final prototype application.}
    \label{fig:proto-zoomout}
\end{figure}

Based on the results of the usability tests, a functional prototype was developed. The result was a web application with a front-end developed with React.JS\footnote{https://react.dev/; Version 18.2.0} and the ChakraUI component library\footnote{https://chakra-ui.com/; Version 2.8.2} and a server developed with Python\footnote{https://www.python.org/; Version 3.12.0} using Flask\footnote{https://flask.palletsprojects.com/en/3.0.x/; Version 3.0.0}. As a final measure, the Aalto Interface Metrics \cite{oulasvirta_aalto_2018} evaluation tool’s UMSI \cite{fosco_predicting_2020} functionality was used to confirm that the interface maintained a coherent visual hierarchy. The user flow in the prototype can be seen in Figure \ref{fig:teaser}, with an overview of the editing and evaluating interface in Figure \ref{fig:proto-zoomout}. The initial wireframe, the resulting prototype, the saliency heatmap and an example sequence diagram of interacting with the application are included in Appendix \ref{app:proto}.

The server employed the DeepL and OpenAI Python libraries\footnote{https://github.com/DeepLcom/deepl-python; Version 1.16.1}\footnote{https://github.com/openai/openai-python; Version 1.3.8} to generate neural machine translations and GPT-4-generated translation evaluations respectively. DeepL was chosen over GPT-4 as the translation provider due to its API providing reliable output and promising results in translating scientific text \cite{takakusagi_validation_2021}.

Two evaluation methods were offered: GEMBA-DA \cite{kocmi_large_2023} (using GPT-4 and the [noref] version of the prompt) and a custom semantic similarity assessment (SSA; using GPT-4-1106-preview for its JSON mode) intended to provide concrete suggestions for improving the semantic similarity of the translation compared to the original (Appendix \ref{app:ssa-prompt}). As the results by Kocmi and Federmann \cite{kocmi_large_2023} and Lu et al. \cite{lu_error_2023} indicated that GPT-based evaluations would not achieve a desired level of quality in segment-level evaluation, the prompt included the original and translated contents in full.

The code for the prototype is publicly available on GitHub\footnote{Front-end: https://github.com/otsha/questionnaires-for-everyone}\footnote{Back-end: https://github.com/otsha/questionnaires-for-everyone-server}.

\section{Study 1: Prototype-assisted Translation of the ATI Scale from English to German}
\subsection{Method}

As discussed, large language models have shown promise in delivering human-readable translation quality evaluations \cite{kocmi_large_2023, lu_error_2023}. While delivering a good user experience was one of the goals of developing the prototype, it was only a means to answer the present study’s research question:

\textit{\textbf{RQ:} Do LLM-generated evaluations help users reach similar questionnaire translation quality to that achieved with conventional, non-NLP-supported, methods?}

The study was conducted online with the prototype hosted on Netlify\footnote{https://www.netlify.com/}. To compare conventional and LLM-assisted translation methods, participants were instructed to utilize the tool to translate the ATI questionnaire \cite{franke_personal_2019} from English to German {\textemdash}, a language a validated version of the questionnaire is available in. Each participant was asked to use the tool until they thought they had achieved a satisfying result, after which they recorded their GEMBA-DA and semantic similarity scores. The participant scores were then compared with scores received by the conventionally translated and validated German version.

Participant experiences with the system as a whole were gathered with the system usability scale (SUS; \cite{brooke_sus_1996}) and their opinion of the quality of the evaluations with the short version of the user experience questionnaire (UEQ-S; \cite{schrepp_design_2017}). Additional measures gathered participants’ opinions of the correspondence of the scores to the translation quality and thoughts on the evaluations. Demographic information such as the participants’ age, gender, occupation, fluency in Finnish and German and prior experiences with questionnaire translation and using LLMs for translation were also requested.

\subsection{Participants}

10 German-speaking participants (5 M; 5 F) between ages 24 and 59 ($M = 32.9$; $SD = 12.1$) were recruited via Prolific. Participants rated their fluency in English and German on scales of 1 (No knowledge at all) to 7 (Native). Median German fluency was 4.0 ($IQR=3.75$; $Min=2.0$; $Max=7.0$) with 1 participant estimating their fluency to be 1{\textendash}2 and 4 participants giving their fluency a rating of 6{\textendash}7. Median English fluency was 6.0 ($IQR=0.75$; $Min=4.0$; $Max=6.0$).

4 participants reported having participated in questionnaire translation before, while 4 had used LLMs for translation in general.

Informed consent was collected from each participant at the beginning of the study. The participants were also informed of their right to stop participating at any time. Participants were compensated £2.00 upon completion of the study, which took them approximately 9 minutes.

\subsection{Analysis}

1 participant was excluded due to not translating the questionnaire as instructed. 2 participants were further excluded due to them not using the designated method for importing their results to the questionnaire. This left 7 participants for the analysis.

In order to be able to compare the translations achieved by the participants to the versions translated using conventional, non-NLP-supported methods, the conventionally translated German language version of the ATI \cite{franke_personal_2019} was evaluated using the two metrics {\textemdash} GEMBA-DA and SSA {\textemdash} available in the tool. Following the number of participants left after exclusions and to account for potential variation in scores given by GPT-4, each metric was run on the original questionnaire 7 times.

\subsection{Results}

\subsubsection{Evaluation Scores}

The original German translation of the ATI scale received a mean GEMBA-DA score of 100 ($SD=0$) and a mean SSA score of 97.86 ($SD=2.47$).

Four participants reported implementing at least some of the suggestions generated by the evaluation. However, only four participants returned their evaluation scores. Of those, all achieved scores of 100 out of 100 for both GEMBA-DA and SSA evaluations. We used the evaluation prompts to separately evaluate the remaining three translations, all of which also scored 100 on both measures.

\subsubsection{Participant Experience}

5 participants reported having implemented at least some of the changes suggested by the semantic similarity evaluation (in response to \textit{"How many of the suggestions given in the evaluation did you implement?"}; 4-step scale: \textit{"None", "Some", "Most", "All"}). When asked to rate how well the numeric evaluation scores matched the translation quality (on a scale from 1{\textendash}7; \textit{Very badly{\textendash}Very well}), participants gave the evaluations a median score of 7.0 ($IQR=0.5$). 

The median SUS score given to the prototype was 92.50 ($IQR=13.75$; $Mean=86.43$; $SD=16.47$), indicating a relatively high degree of usability. The UEQ-S scores indicate that participants perceived the prototype to have a higher pragmatic ($Median=1.50$; $IQR=0.38$) than hedonic ($Median=0.50$; $IQR=1.63$) quality.

One participant was particularly impressed by the evaluation quality: \textit{"I'm surprised at how accurate the evaluation is. Although not a professional translator, I have done some translation work in the past, and I think this tool can further enhance the quality of translations in the context of questionnaires."} (P10). Another might have been slightly confused by GPT-4's tendency to suggest changes even if none were needed: \textit{"I think the evaluations made were pretty spot on. Some just said it were not matching exactly the english version but from what I could tell it seemed exactly the same to me"} (P8).

\section{Study 2: Prototype-assisted Translation of the BFI-10 Scale from English to Portuguese}

In Study 1, Participants managed to achieve evaluation scores similar to those received by the conventionally translated questionnaire. Outlook on the user experience of the tool was generally positive.

The most significant limitations of this study were the low number of participants and that participants did not follow the instructions correctly. Due to these issues, we decided to run a follow-up study with a larger sample that would clarify the instructions.

\subsection{Method}

The follow-up study was shorter and focused more on the evaluations. As the user experience of the tool seemed to be quite good based on the first study, we decided not to include the SUS and UEQ-S questionnaires.

In the second study, participants were tasked to translate the BFI-10 \cite{rammstedt_measuring_2007} personality questionnaire from English to Portuguese; a language the questionnaire has not yet been translated to using conventional methods and one that features a large participant pool on Prolific.

\subsection{Participants}

20 Portuguese-speaking participants (9 M; 11 F) between ages 21 and 44 ($M=27.65$; $SD=6.05$) and no self-reported difficulty with reading were recruited via Prolific. Median self-reported Portuguese fluency was 7.0 ($IQR=0.0$; $Min=5.0$; $Max=7.0$), and English fluency 6.0 ($IQR=1.0$; $Min=4.0$; $Max=6.0$). 6 of the participants had been involved in translating a questionnaire. 12 participants had used LLMs for translation.

Informed consent was collected from the participants at the beginning of the study. Participants were compensated £1.32 upon completion of the study. Based on the first study, we estimated the shortened survey to take the participants approximately 8 minutes, but participants ended up spending more time on completion (around 11 minutes on average).

\subsection{Analysis}

We scored 4 conventionally translated versions of the BFI-10 (Danish \cite{palsson_crosscultural_2023}, French \cite{courtois_validation_2020}, German \cite{rammstedt_measuring_2007} and Chinese \cite{carciofo_psychometric_2016}) individually using both GEMBA-DA and SSA to obtain a baseline for comparison. Matching the number of participants, each version was scored 20 times on both metrics. Then, the scores achieved by the participants using our prototype were compared against scores received by the conventionally translated versions.

\subsection{Results}

\subsubsection{Evaluation Scores}

The conventionally translated and validated versions received a mean GEMBA-DA score of 97.56 ($SD=2.66$) and a mean SSA score of 94.88 ($SD=0.78$). A one-way ANOVA revealed that there was statistically significant language-dependent variation in the mean GEMBA scores received by these versions ($F\textsubscript{Trad-GEMBA-DA}(3, 76) = 17.39; p < 0.001$). Post-hoc pairwise \textit{t}-tests showed that the Chinese-language translation \cite{carciofo_psychometric_2016} was an outlier, receiving lower GEMBA-DA scores ($M\textsubscript{Trad-ZH-GEMBA-DA} = 94.75; SD = 2.86$) compared to the Danish ($t = -4.43; p < 0.001$), French ($t = -6.46; p < 0.001$) and German ($t = -6.01; p < 0.001$) versions. However, no significant language-dependent variation was found in the SSA scores achieved by the conventionally translated versions ($F\textsubscript{Trad-SSA}(3, 76) = 6.0; p > 0.1$). This indicates that even if the translation quality of the Chinese-language version was not as high as the other versions, it delivered the semantics of the original questionnaire equally well.

Translations generated by the participants had a mean GEMBA-DA score of 99.75 ($SD=1.09$; $Min=95$; $Max=100$) and a mean SSA score of 95.75 ($SD=2.75$; $Min=90$; $Max=100$). The tool-assisted GEMBA-DA scores did significantly differ from scores achieved by conventional methods ($F\textsubscript{GEMBA}(4, 95)=20.27; p < 0.001$). Pairwise \textit{t}-tests showed that participants achieved significantly higher scores than both the Danish ($t = 3.33; p < 0.01$) and Chinese ($t = 8.12; p < 0.001$) versions. The SSA scores achieved by participants did not significantly differ from those received by the conventionally translated versions ($F\textsubscript{SSA}(4, 95)=1.93; p > 0.1$).

The mean scores received by each conventional translation, as well as the mean scores participants achieved, can be found in Table \ref{tab:res-means}.

\begin{table}
    \centering
    \begin{tabular}{|l|c|c|c|c|c|} \hline 
 Translation method& \multicolumn{4}{|c|}{Conventional}&Prototype-assisted\\ \hline 
         Target language&  Chinese&  Danish&  French&  German& Portuguese\\ \hline 
 Mean GEMBA-DA[noref] score& 94.75*& 97.70& 99.05& 98.75&99.75*\\ \hline 
 Mean SSA score& 94.50& 95.00& 95.00& 95.00&95.75\\ \hline
    \end{tabular}
    \caption{The mean scores achieved by each of the conventionally translated versions, as well as the prototype-assisted translations of the BFI-10 \cite{rammstedt_measuring_2007}. The Chinese-language version received significantly lower GEMBA scores than all other versions, while prototype-assisted Portuguese translations received significantly higher scores than the Chinese and Danish versions.}
    \label{tab:res-means}
\end{table}

\subsubsection{Participant Experience}

Participants found the suggestion quality high ($Median=6.0$; $IQR=1.0$). While 12 participants reported having implemented at least some of the suggestions, only 1 reported having implemented most, with no participants having implemented all of them. Nearly all (18) participants reported having manually edited the translations.

One participant was not quite satisfied with the quality of the evaluation: \textit{"[...] I had to manually edit the second item of the questionnaire to preserve it's original meaning [...] The tool suggested to revert to the original translation [...] which has a different meaning"} (P19). Another noted that as the initial translation had received a perfect score, they did not receive any suggestions to implement: \textit{"Since no suggestions for correction were made, I can not tell how useful this feature is."} (P18).

\section{General Discussion and Limitations}

The results from our two studies show promising results regarding LLM adoption in the questionnaire translation process. We found that machine translation, accompanied by GPT-4 generated quality scores, helped participants reach at least a similar level of translation quality and semantic similarity to questionnaire translations generated using conventional, non-NLP-supported methods. Participants found the evaluations to accurately represent translation quality.

Participants found the suggestions provided by our experimental semantic similarity evaluation to be moderately helpful. In the second study, while nearly all participants had opted to edit the translations manually, only just over half reported having implemented suggestions. A factor limiting meaningful evaluation of how practical the suggestions provided by the tool was the high performance of the DeepL translator. As the initial translations were scored highly, the suggestions generated by GPT-4 were of limited value.

One factor for the high scores received by the translations generated by participants could be the high initial quality of the questionnaires. As noted by \cite{brislin_back-translation_1970, su_generating_2002}, the wording of the original questionnaire highly affects how well its semantics translate to other languages. Future research should investigate LLM translation quality evaluations with questionnaires that have been found more difficult to translate due to culture-specific expressions.

Another factor contributing to high translation quality could have been the close relationship of the languages. English, German and Portuguese are all Indo-European languages, with English and German especially close: both are Germanic languages \cite{kapovic_indo-european_2017}. Another topic for future research is investigating LLM assessments when questionnaires are translated between language families.

Beyond making the translation of questionnaires more accessible, LLMs could be used for their preliminary validation as well. Schmidt et al. \cite{schmidt_simulating_2024} provide an overview of the potential use cases of LLMs in complementing human participants in human-centric design. Studies by Tavast et al. \cite{tavast_language_2022} and Hämäläinen et al. \cite{hamalainen_evaluating_2023} have found that LLMs can be used to simulate participant responses in both Likert-style questionnaires and open-ended questions. Future research should, therefore, investigate whether LLMs can be prompted to generate questionnaire validation results similar to those obtained through user testing with real people.

We provide two design recommendations regarding the integration of LLM-generated evaluations into translation software. First, we noticed that evaluation generation with LLMs can take time, especially when dealing with longer questionnaires. To avoid user frustration, it is important to communicate longer wait times to the user in some manner \cite{nielsen_enhancing_1994}; we simply used a spinner accompanied by a short message. Second, to avoid evaluations or suggestions that might be perceived as redundant, prompt engineering could discourage the system from providing feedback after a certain score threshold.

\section{Conclusion}

LLM-generated translation quality evaluations can help researchers identify and resolve issues in their translations based on context-specific requirements. This work presents promising results regarding the integration of LLM-generated feedback to translation software to expedite the questionnaire translation process for cross-cultural research. We demonstrate that GPT-4-generated translation quality assessments, alongside machine translation, can help users independently reach translation results similar to those attained with conventional, non-NLP-supported methods.

It should be kept in mind that validation is as much an integral part of the questionnaire adaptation process as translation. Questionnaires, whether translated using automated tools or conventional methods, should be validated with participants before deploying them in research.



\bibliographystyle{ACM-Reference-Format}
\bibliography{translator-proj}

\clearpage

\appendix

\section{Task Analysis Diagrams}\label{app:hta}

\begin{figure}[h]
    \centering
    \includegraphics[width=\textwidth]{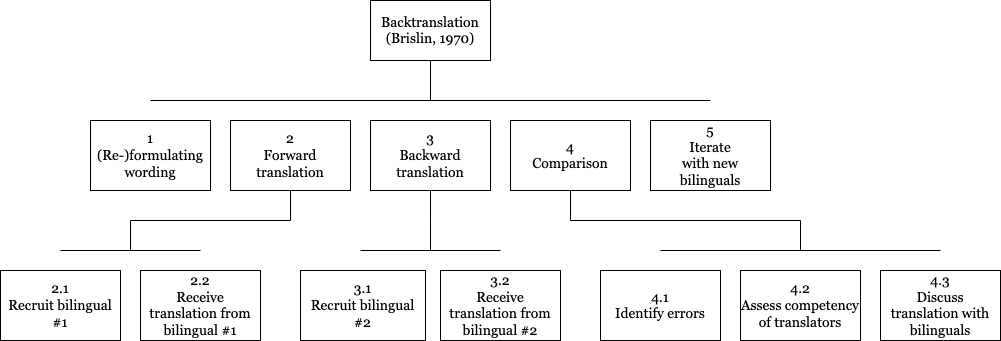}
    \caption{Brislin's \cite{brislin_back-translation_1970} backtranslation approach}
    \label{fig:bristlin-bt-hta}
\end{figure}

\begin{figure}[h]
    \centering
    \includegraphics[width=\textwidth]{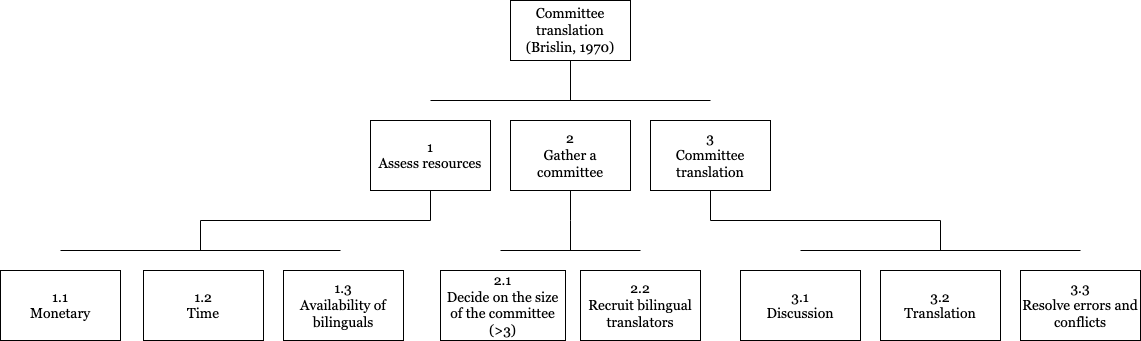}
    \caption{Brislin's \cite{brislin_back-translation_1970} committee approach}
    \label{fig:bristlin-committee-hta}
\end{figure}

\begin{figure}[h]
    \centering
    \includegraphics[width=\textwidth]{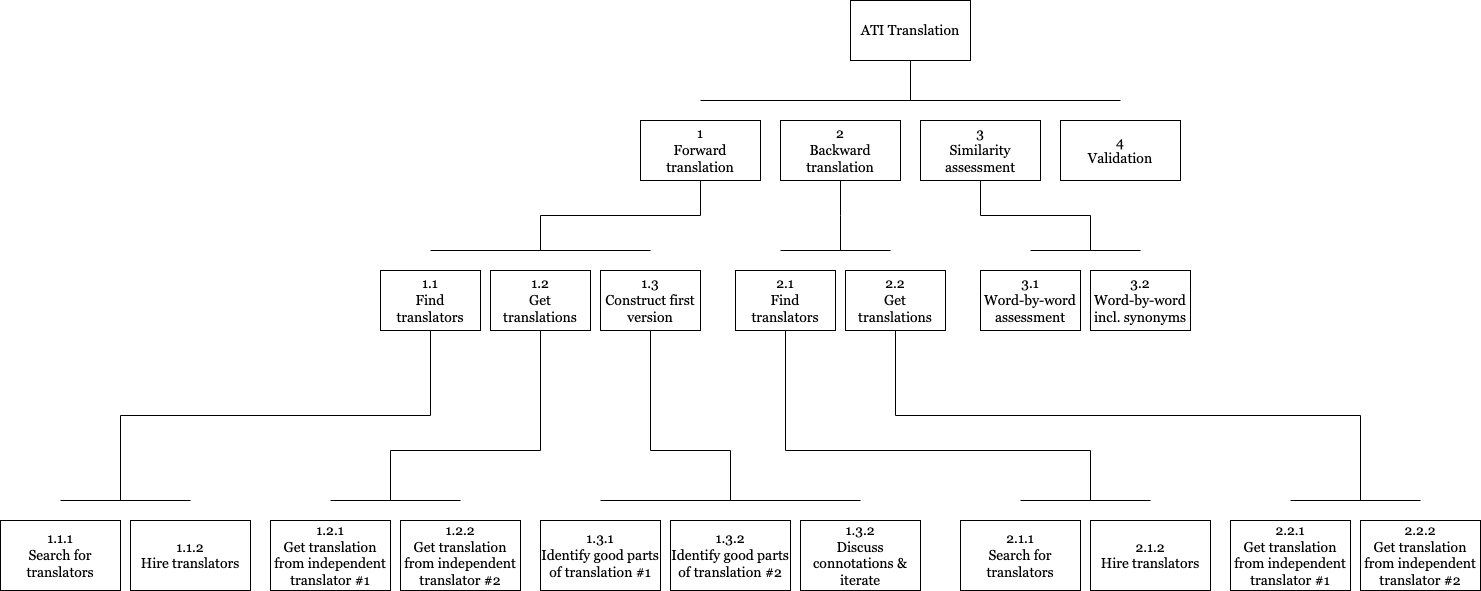}
    \caption{Finnish ATI translation process \cite{heilala_finnish_2023}}
    \label{fig:fi-ati-hta}
\end{figure}

\begin{figure}[h]
    \centering
    \includegraphics[width=\textwidth]{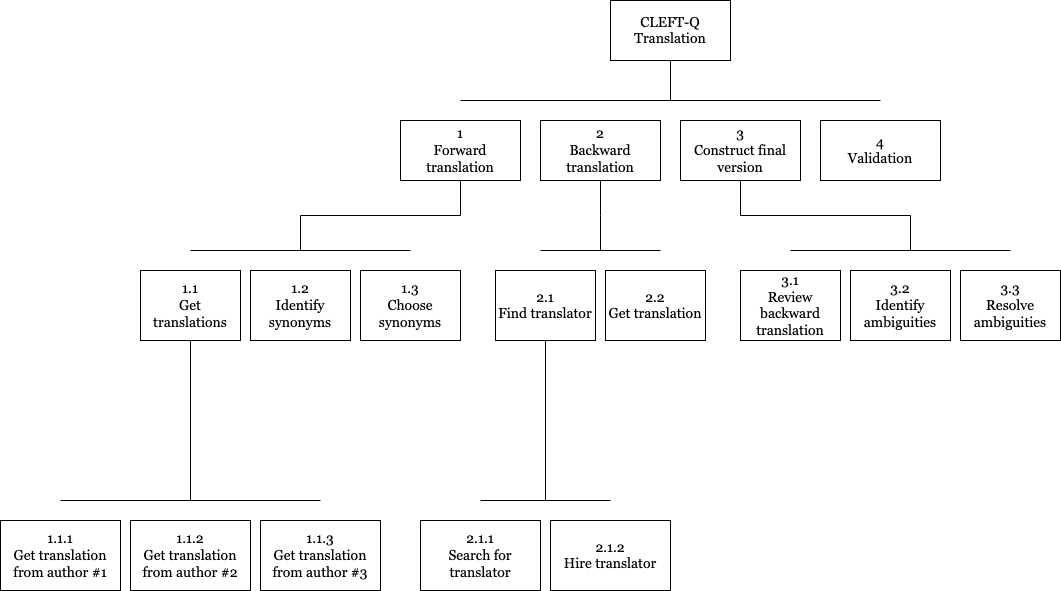}
    \caption{Finnish CLEFT-Q translation process \cite{westerlund_finnish_2023}}
    \label{fig:fi-cleftq-hta}
\end{figure}

\clearpage

\section{Prototyping}\label{app:proto}

\subsection{Initial High-fidelity Prototype}

\begin{figure}[h]
    \centering
    \includegraphics[width=\textwidth]{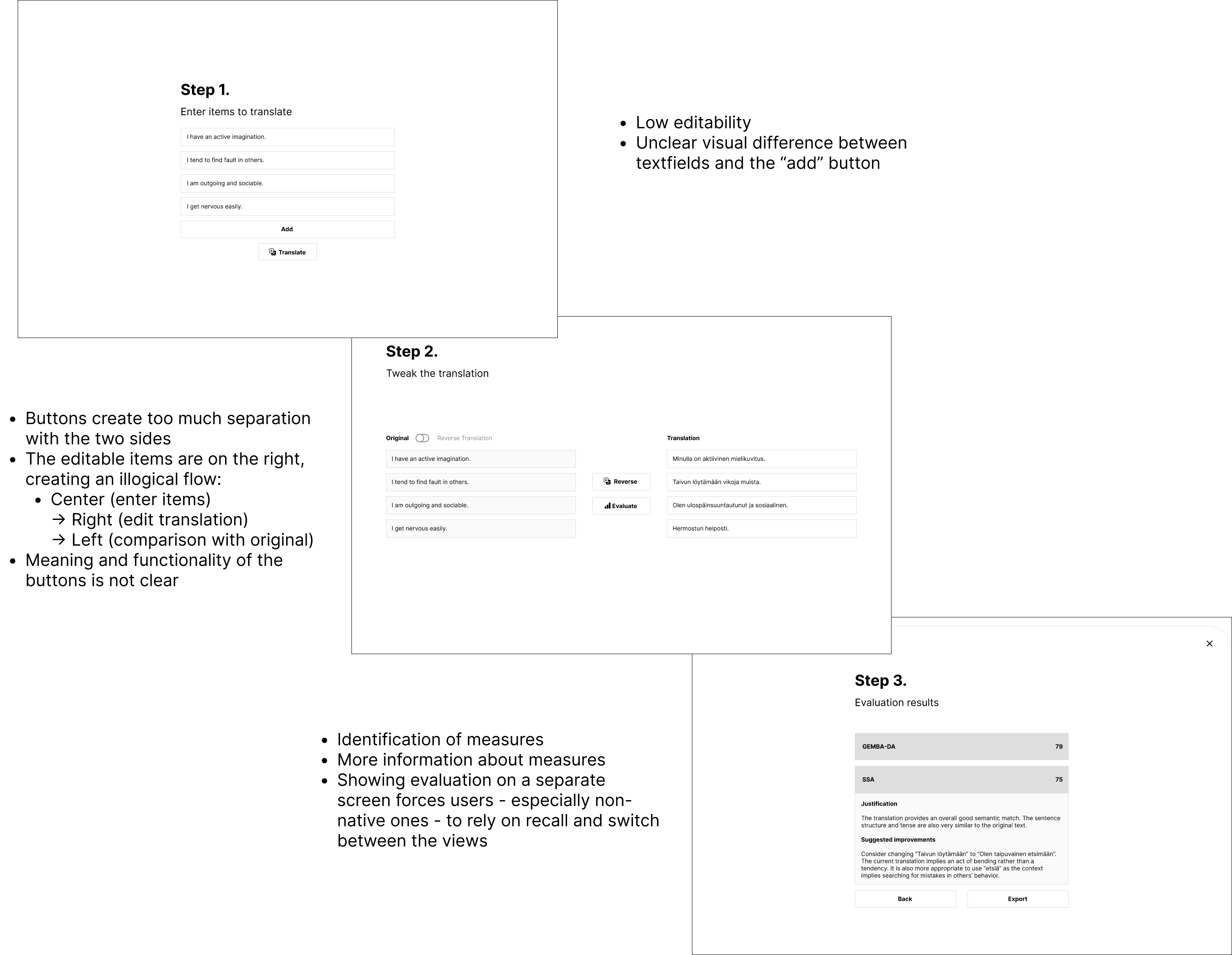}
    \caption{The initial high-fidelity non-functional interactive prototype that was used for usability testing. The  actions taken included changing item labeling and addition of tooltips, reordering of the layout to match user expectations of left-to-right logical flow, moving the buttons away from in between the sides to bring the two sides closer together and delivering the evaluations on the same screen for less reliance on recall.}
    \label{fig:initial-proto}
\end{figure}

\clearpage

\subsection{Final Prototype}

\begin{figure}[h!]
    \centering
    \includegraphics[width=\textwidth]{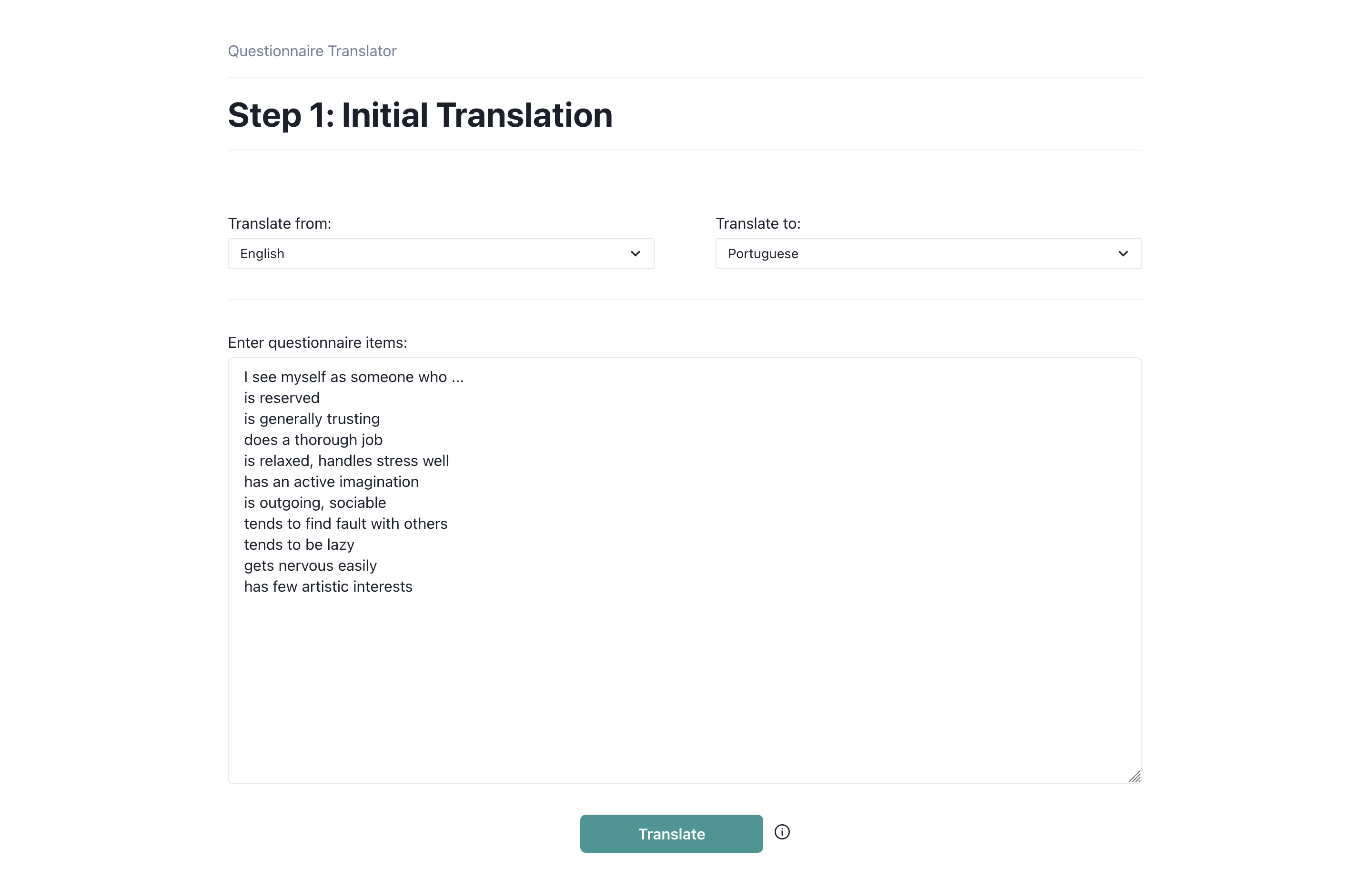}
    \caption{The initial view of the prototype translation application features language selection and a textarea for inputting the questionnaire items.}
    \label{fig:final-proto-step1}
\end{figure}

\begin{figure}[h!]
    \centering
    \includegraphics[width=0.8\textwidth]{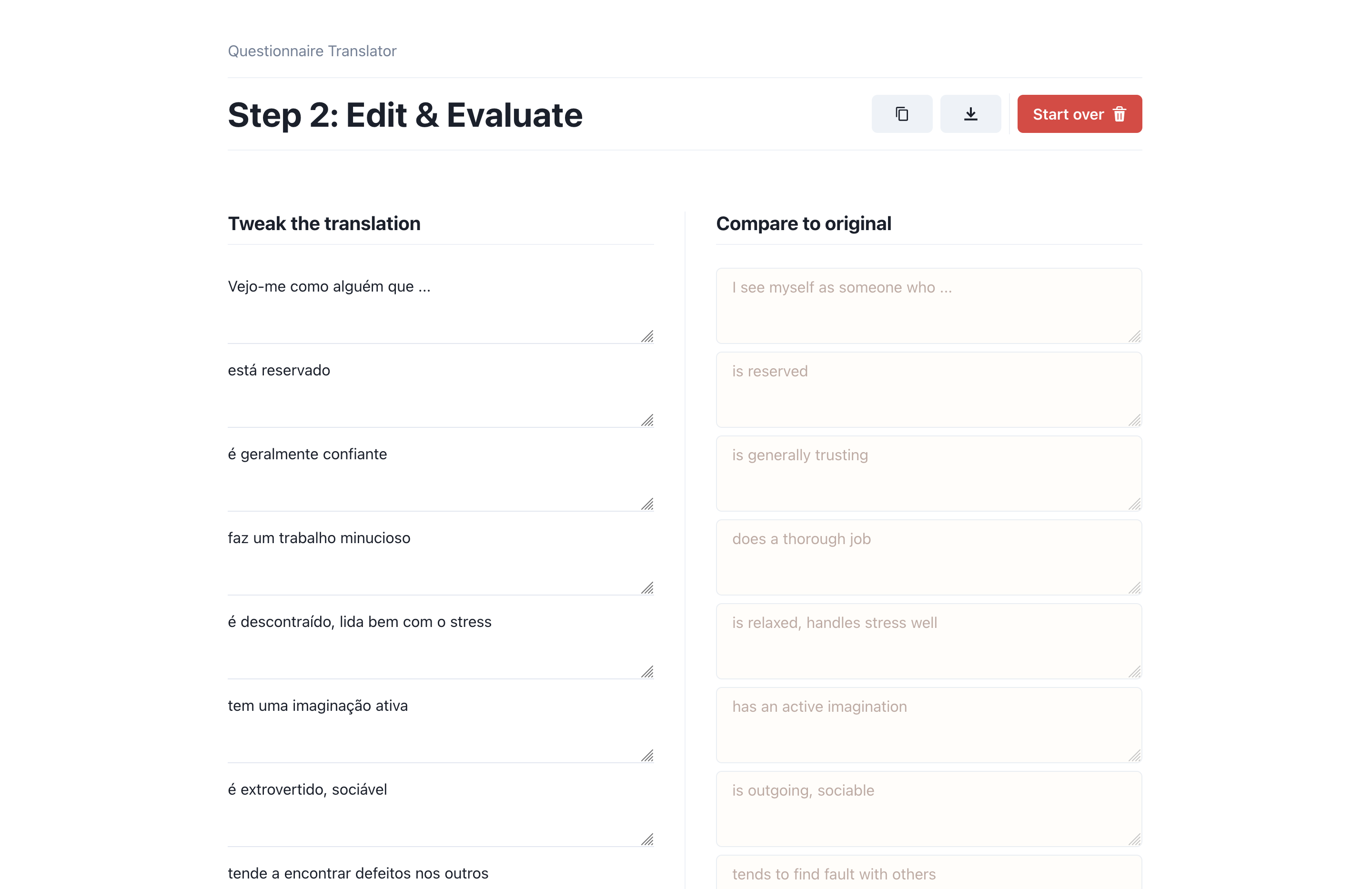}
    \caption{The view after performing the initial translation. On the left: The initial item translations, which can be edited manually by the user. On the right, the original items.}
    \label{fig:final-proto-step2a}
\end{figure}

\begin{figure}[h!]
    \centering
    \includegraphics[width=0.8\textwidth]{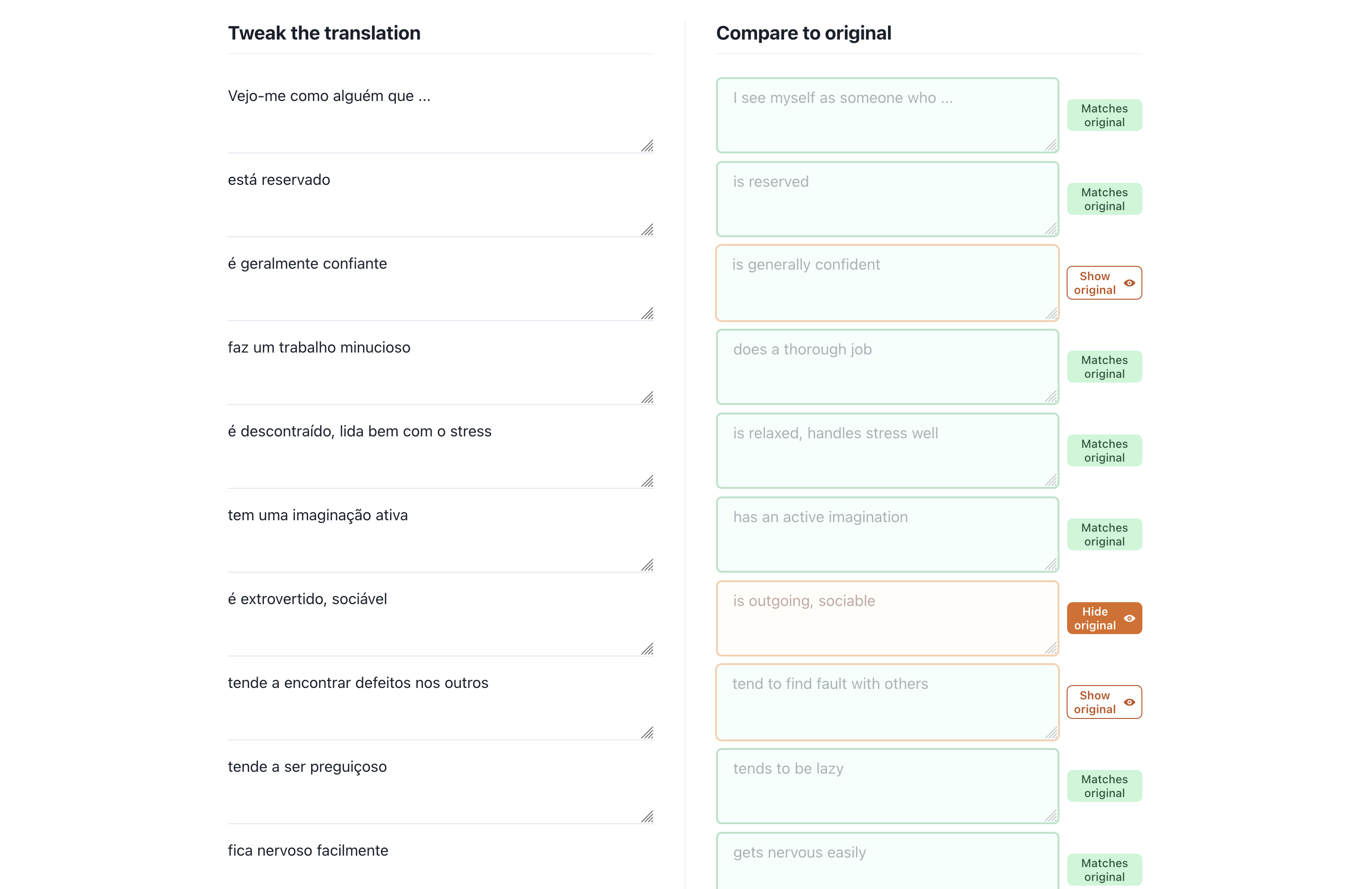}
    \caption{The view after conducting a backtranslation. The right side allows the users to switch between the original and the backtranslated version of each non-matching item.}
    \label{fig:final-proto-step2b}
\end{figure}

\begin{figure}[h!]
    \centering
    \includegraphics[width=\textwidth]{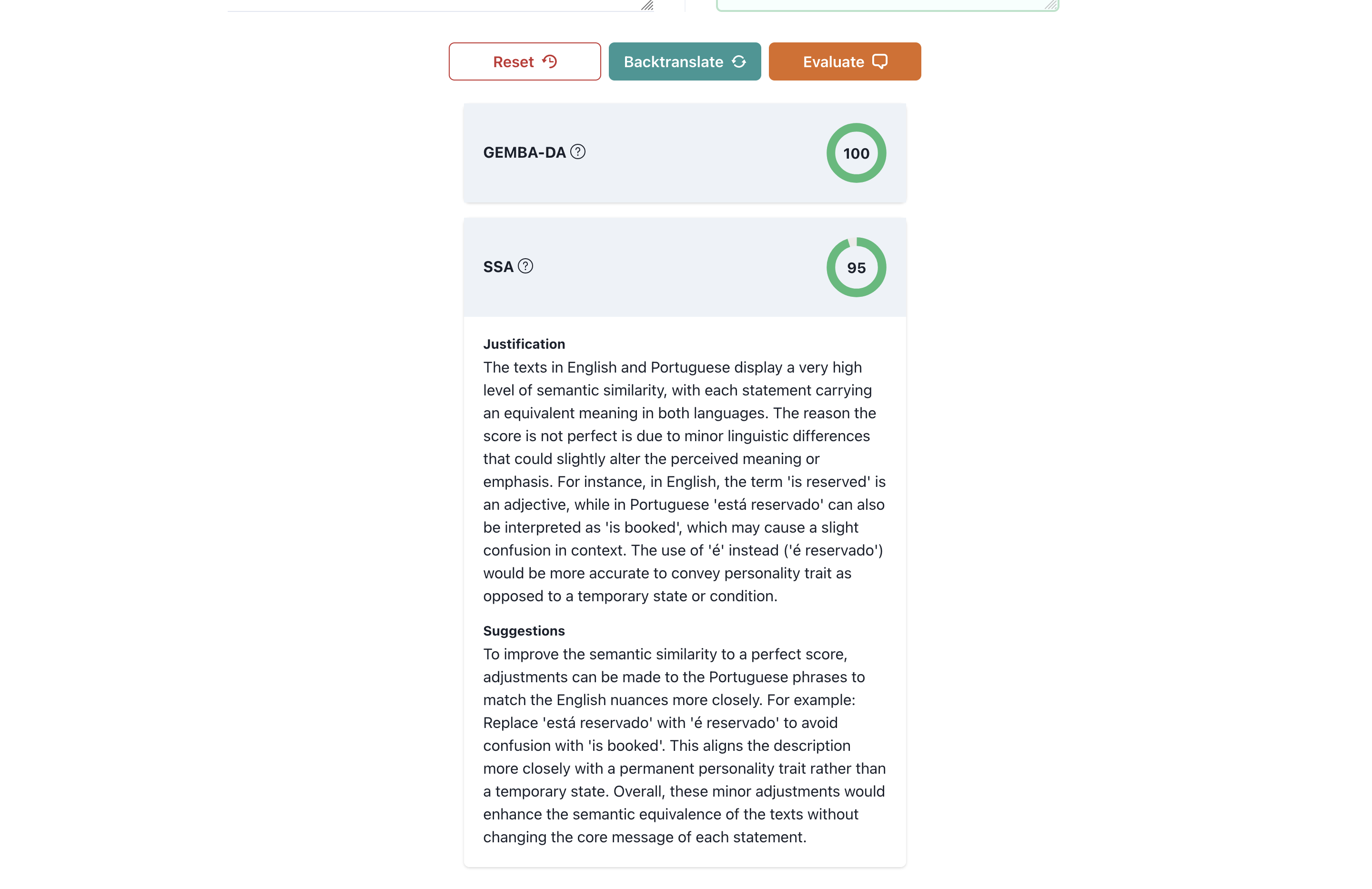}
    \caption{The evaluations provided by the application. The semantic similarity assessment gives users a verbal justification of the score given, coupled with suggestions for improvement.}
    \label{fig:final-proto-step2c}
\end{figure}

\begin{figure}[h!]
    \centering
    \includegraphics[width=\textwidth]{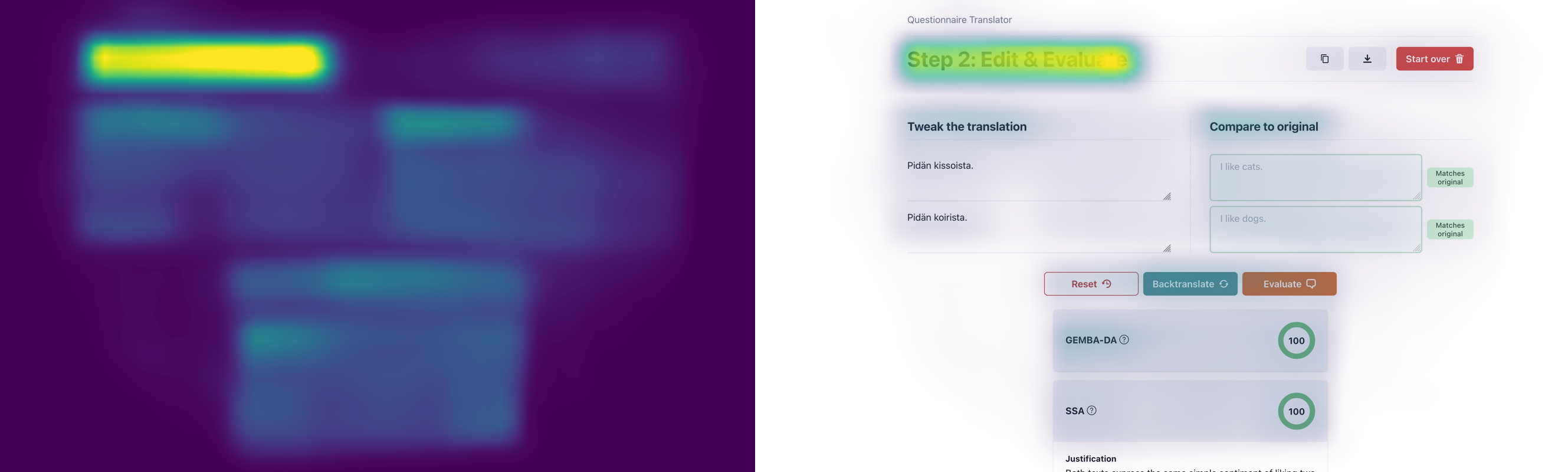}
    \caption{UMSI saliency evaluation of the interface using the Aalto Interface Metrics service \cite{oulasvirta_aalto_2018}. The instructional texts attracted the most emphasis, followed by the core actions of backtranslation and evaluation. The top-right toolbar was the least salient.}
    \label{fig:umsi-saliency}
\end{figure}

\begin{figure}[h!]
    \centering
    \includegraphics[width=\textwidth]{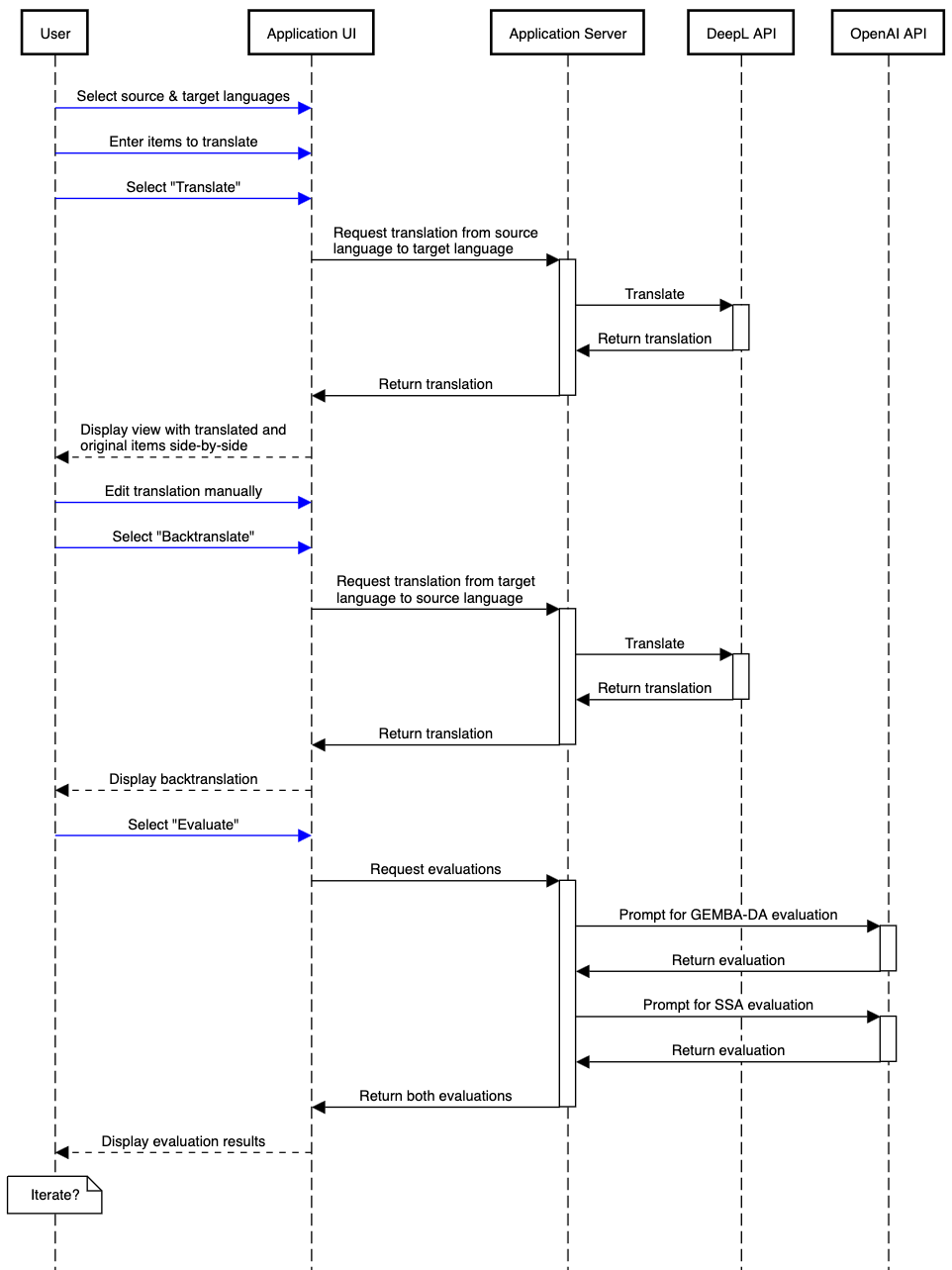}
    \caption{An example sequence diagram of the process of using the application.}
    \label{fig:tlt-app-seq}
\end{figure}

\clearpage

\section{Prompt for Semantic Similarity Assessment}\label{app:ssa-prompt}

Assess the semantic similarity of the following texts in \{source\_lang\} and \{target\_lang\} on a scale from 0 (no semantic similarity at all) to 100 (perfect semantic similarity). Justify the score. Provide a single paragraph suggesting changes to the \{target\_lang\} version (i.e. word or expression replacements) to improve the score.
\\ \\
\noindent \{source\_lang\}: "\{source\_text\}"
\\
\noindent \{target\_lang\}: "\{target\_text\}"
\\ \\
\noindent Respond with JSON only in the following format:
\\ \\
\noindent score
\\
\noindent reasoning,
\\
\noindent suggestion

\clearpage

\end{document}